# Stabilization of interfaces for double-cation halide perovskites with $AVA_2FAPb_2I_7$ additives

Lev O. Luchnikov[1], Ekaterina A. Ilicheva[1], Victor A. Voronov[1], Prokhor A. Alekseev[2,3], Mikhail S. Dunaevskiy[2,3], Vladislav Kalinichenko[2], Vladimir Ivanov[2], Aleksandra Furasova[2,4], Daria A. Krupanova[5], Ekaterina V. Tekshina[5], Sergey A. Kozyukhin[5], Dmitry S. Muratov[6], Maria I. Voronova[1], Danila S. Saranin[1,2,#], and Eugene I. Terukov[2]

**Affiliations:**

[1]LASE – Laboratory of Advanced Solar Energy, NUST MISiS, 119049 Moscow, Russia

[2]School of Physics and Engineering, ITMO University, 197101, St. Petersburg, Russia

[3]Ioffe Institute, 194021, St. Petersburg, Russia

[4]Qingdao Innovation and Development Center, Harbin Engineering University, Qingdao 266000, Shandong, China

[5]Kurnakov Institute of General and Inorganic Chemistry of the Russian Academy of Sciences, Moscow, 119991, Russia

[6]Department of Chemistry, University of Turin, 10125, Turin, Italy

[#]*Corresponding authors*

**Corresponding authors:**

saranin.ds@misis.ru - Dr. Danila S. Saranin

**Abstract:**

The use of mixed cation absorber composition was considered as an efficient strategy to mitigate the degradation effects in halide perovskite solar cells. Despite the reports about partial stabilization at elevated temperatures, unfavorable phase transition after thermocycling and electric field-driven corrosion remains critical bottlenecks of perovskite thin-films semiconductors. In this work, we developed stabilized heterostructures based on $CsFAPbI_3$, modified with mechanically synthesized quasi-2D perovskite incorporating the 5-ammonium valeric acid cation ($AVA_2FAPb_2I_7$). We found that integration of $AVA_2FAPb_2I_7$ into grain boundaries boosts phase resilience under harsh thermocycling from -10 up to 100°C and suppresses transitions, as well as decomposition to $PbI_2$. The rapid oxidation of metal contacts in the multi-layer stacks with non-passivated $CsFAPbI_3$ was effectively suppressed in the fabricated heterostructure. A comprehensive interface study of the copper electrode contact revealed that the incorporation of $AVA_2FAPb_2I_7$ stabilized the lead and iodine states and suppressed contamination of FA cation in ambient conditions. Meanwhile, the metal/perovskite interface remained predominantly in the Cu(0)/Cu(I) state. The observed stabilization in perovskite heterostructure was attributed to an increased activation energy for δ-phase accumulation at the grain boundaries combined with reduced ionic diffusion. The obtained results opened important highlights for the mechanisms of the improved phase stability after thermal cycling and mitigation of the interface corrosion and under an applied electric field.

**Introduction:**

Halide perovskite photovoltaics (HP-PV) represents one of the most promising third-generation solar cell technologies[1], [2]. Currently, perovskite solar cells (PSCs) have reached the record power conversion efficiency (PCE) of 26.7%[3], which is comparable to crystalline silicon and exceeding the performance of alternative PV technologies such as CIGS (copper-indium-gallium-selenide), CdTe, and OPV (organic photovoltaics)[4], [5], [6]. Solution processing methods like slot-die coating and inkjet printing could enable cost-effective scaling from cells to modules, reducing the capital expenditure (CAPEX) needed for mass production[7]. By 2024, over eight pilot plants are under development for high-throughput fabrication of mega-watt scale. To meet industry standards, solar cells must demonstrate extended operational lifetimes and minimal efficiency degradation during the exploitation period. Rapid corrosion at the interfaces and the decomposition of the perovskite molecule under high-intensity light, thermal cycling, and moisture remain critical obstacles to achieving reliable device stabilization[8].

Degradation in thin-film perovskite device structures at the elevated temperatures typically involves the formation of hydrogen iodide (HI), molecular iodine ($I_2$), gaseous decomposition products, lead iodide, and the transition to non-photoactive phases (e.g., $\delta$-$CsPbI_3$ and $\delta$-$CsFAPbI_3$)[9]. The substitution of methylammonium with formamidinium and inorganic cesium in multi-cationic compositions has led to enhanced thermal stability of the photoactive phase[10]. However, corrosion effects at the absorber/charge-transport layer interfaces are also evident in the modified compositions. Microcrystalline thin-films of halide perovskites are characterized by numerous uncompensated defects, such as iodine interstitials, vacancies, and anti-sites[11], [12]. The migration of the ionic defects also adversely affects metal electrodes. It is well established that the perovskite/metal contact is prone to immediate chemical interactions, resulting in instability[13], [14], [15]. However, in device structures, these effects are partially mitigated by the inclusion of charge-transport layers. Prolonged exposure of PSCs with standard back electrodes (Ag, Au, Cu) to exploitation conditions (temperature 25 - 85°C, light intensity ~100 mW/cm$^2$) leads to electrode oxidation, delamination, and a reduction in photoelectric performance[16], [17].

Among the others approaches, the use of perovskite heterostructures was considered as an effective strategy to mitigate the corrosion effects at the interfaces of PSCs. The heterostructure of halide perovskites typically consists of a standard microcrystalline absorber film combined with low-dimensional perovskite configurations. Replacing the A-site cation with a large spacer molecule (typically phenyl-ethyl ammonium (PEA) or $C_4H_9NH_3$ (BA)) induces spatial separation within the 3D perovskite crystal lattice, resulting in alternating layers of 3D perovskite and large organic spacers[18]. The molar ratio between the spacer molecule and the perovskite framework dictated the dimensionality of the resulting phase. This results in the formation of so called quasi-two-dimensional (2D) Ruddlesden-Popper phases coupled with the lattice of the photoactive three-dimensional (3D) perovskite phase. Current literature[19], [20], [21], [22] highlights the effects of surface passivation in perovskite films, including reduced defect concentrations at inter-grain interfaces and improved thermal stability in 2D/3D perovskite heterostructures (**PHS**). The integration methods of 2D perovskite additives also strongly impacts the crystal structure of perovskite thin-films. A pioneering study by *Loi group* demonstrated[23] the impact of employing a 2D perovskite template, deposited via a scalable technique, which subsequently undergoes in situ conversion to form a highly crystalline 3D perovskite. The authors revealed that 2D perovskite integration rules the grain orientation and reduces the intergranular boundary surface.

Interface stabilization in PSCs presents a challenging task, as it involves the analysis of the absorber, charge-transport layers, and thin-film electrodes. Ionic defect migration contributes significantly, particularly to device structures under applied electric fields[24], [25]. Many studies on PSCs primarily focus on the properties of PHS - absorbers and photo-carrier transport, however, a comprehensive understanding of stabilization mechanisms must also consider the state of thin-films and corrosion at interfaces. The use of multi-cationic perovskite absorbers is a well-established strategy for achieving both high power conversion efficiency (PCE) and operational stability exceeding thousands of hours[26], [27]. While 2D perovskites and template-assisted growth have been extensively studied for single-cation compositions, the development of heterostructures with multi-cationic perovskites necessitates precise stoichiometric tuning. It is also crucial to account for potential phase segregation effects induced by heating and phase transitions during thermal cycling. Despite promising reports on partial stabilization of perovskite devices under elevated temperatures and damp-heat conditions[28], a deeper understanding of passivation mechanisms is required to achieve degradation rates comparable to silicon[29] and CdTe[30] photovoltaics (<1% per year of exploitation[31], [32]). The challenges of enhancing phase stability, improving surface passivation, and mitigating ionic effects remain critical.

In this work, we present a comprehensive study on the chemical stability of perovskite heterostructures (PHS) based on microcrystalline $CsFAPbI_3$ and mechano-synthesized $AVA_2IPb_2I_7$, fabricated via vacuum-assisted solution processing (VASP). To assess the impact of the electric field effects on ion defect - induced corrosion, we investigated thin-films of PHS as well as multilayer stacks with electrodes. The investigation included potential probing of the inter-grain areas, phase composition changes in exposure to external stress factors - moisture, oxygen, and thermal cycling across a temperature range of -10 to 100°C. We found that the PHS and bare microcrystalline perovskite showed distinct corrosion behaviors, with the PHS displaying enhanced durability against phase segregation. Complex investigation of the surface properties revealed the essential role of $AVA_2IPb_2I_7$ for the stabilization of perovskite thin-films with metal contacts, compensating the states of iodine, lead and nitrogen even in presence of ambient humidity and oxygen. The results obtained were deeply analyzed and discussed.

## Results and discussions

For the fabrication of HP- based thin-films, we applied vacuum-assisted solution processing (**VASP**). The principle of VASP involves applying reduced pressure to the wet film after depositing the perovskite solution on the substrate (**fig.1(a)**). This process facilitates controlled solvent removal and saturates the solution within the wet film, providing optimal conditions for nucleation. Compared to the standard solvent-engineering approaches for crystallizing perovskite absorbers, VASP is a promising technique for up-scaling module fabrication due to improved homogeneity of the crystallization process across the large substrates[33], [34]. Two-cation $Cs_{0.2}FA_{0.8}PbI_{2.93}Cl_{0.07}$ (CsFAPbI$_3$) was used as a reference microcrystalline absorber. Perovskite heterostructures were formed with a quasi-2D $AVA_2FAPb_2I_7$, which is structured around the spacer molecule of 5-AVAI (5-Ammonium valeric acid iodide). $AVA_2FAPb_2I_7$ was mechanically synthesized in a planetary mill using a stoichiometric composition of the precursors (**Fig.1 (b)**, see details in electronic supplementary material (**ESI**)). The X-ray diffraction (**XRD**) pattern of the $AVA_2FAPb_2I_7$ after ball milling exhibits peaks at 13.92° and 28.07°, corresponding to the 110 and 220 planes of tetragonal P4/mbm β-CsFAPbI3, respectively. Peaks observed at 4.66° and 5.79° are associated with planes of the perovskite material, separated by the 5-AVAI spacer molecule. This confirms the formation of a quasi-2D perovskite structure during the mechanoactivation process. The solution precursors for VASP fabrication were made in a mixture of dimethyl formamide (**DMF**) and N-methyl pyrrolidine (**NMP**). To form the PHS, the powder of mechano-synthesized $AVA_2FAPb_2I_7$ was added to the solution of CsFAPbI$_3$ in concentration of 1.0 mol%. The pre-synthesized $AVA_2FAPb_2I_7$ promoted the formation of two-dimensional perovskite clusters, which acted as crystallization centers for the heterostructure. To simplify sample identification, we used the following names: "Control' for bare CsFAPbI$_3$ and "Target" for PHS with $AVA_2FAPb_2I_7$.

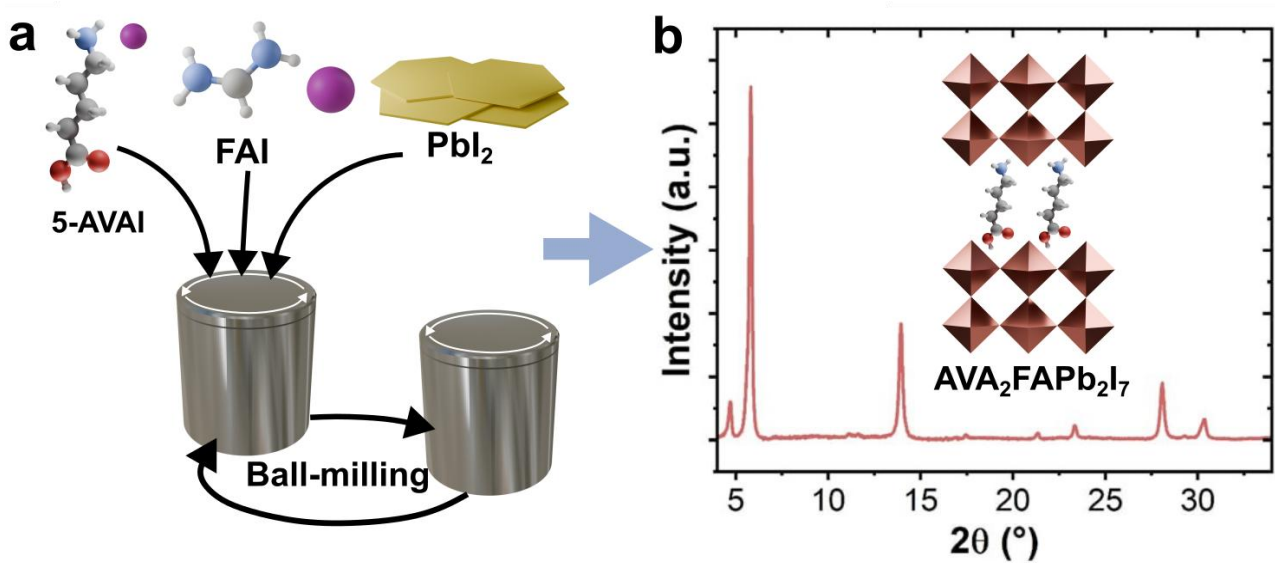

Figure 1 – General schematics of the mechanical synthesis for AVAFAPb$_2$I$_7$ (a) sketch of molecular schematics for $AVA_2FAPb_2I_7$ and corresponding X-ray diffractogram for the synthesized powder

The changes of surface morphology were estimated using atomic force microscopy (AFM) simultaneously with Kelvin probe force microscopy (KPFM). Both sample configurations showed the average grain size ~80 nm, with an RMS roughness of 9–10 nm. Notably, that surface profile of the control films exhibited the inter-grain voids 10–20 nm (**Fig.S1(a)** in **ESI**). For the Target film we observed improved planarity specifically for grain-to-grain interfaces (**Fig.S1(b)** in **ESI**). Areas on control sample are characterized by a decrease in surface potential (**Fig. S1 (c)** in **ESI**). The control film is also uniformly covered with point defects, which correspond to a lower surface potential. In contrast, the surface potential of the target perovskite sample is uniform (**Fig. S1 (d)** in **ESI**). Consequently, the grain structures of the control and target samples are nearly identical, apart from the "dendrite"-like structures present in the control.

We assumed that $AVA_2FAPb_2I_7$ perovskite is distributed at the inter-grain interfaces of $CsFAPbI_3$. To define the changes in the optoelectronic properties of the PHS surface, we employed KPFM to measure the surface potential with more details and precision (**fig.2**). Figure 2 displays the surface potential distribution for control (a) and target (b) samples. **Fig.S1(e),(f)** in **ESI** presents the corresponding topography images. While the topography images exhibit similar relief of target and control samples, the surface potential distributions show a distinct difference in grain boundaries. To highlight the difference, **fig. 2(c)** and **(d)** show corresponding profiles, taken across the boundaries (black dashed lines). From the profiles, it follows 3 times larger potential drop on the boundary in the target sample. One can propose that the increasing of the potential drop is related to the formation of the $AVA_2FAPb_2I_7$ inclusions on the boundary in the target sample.

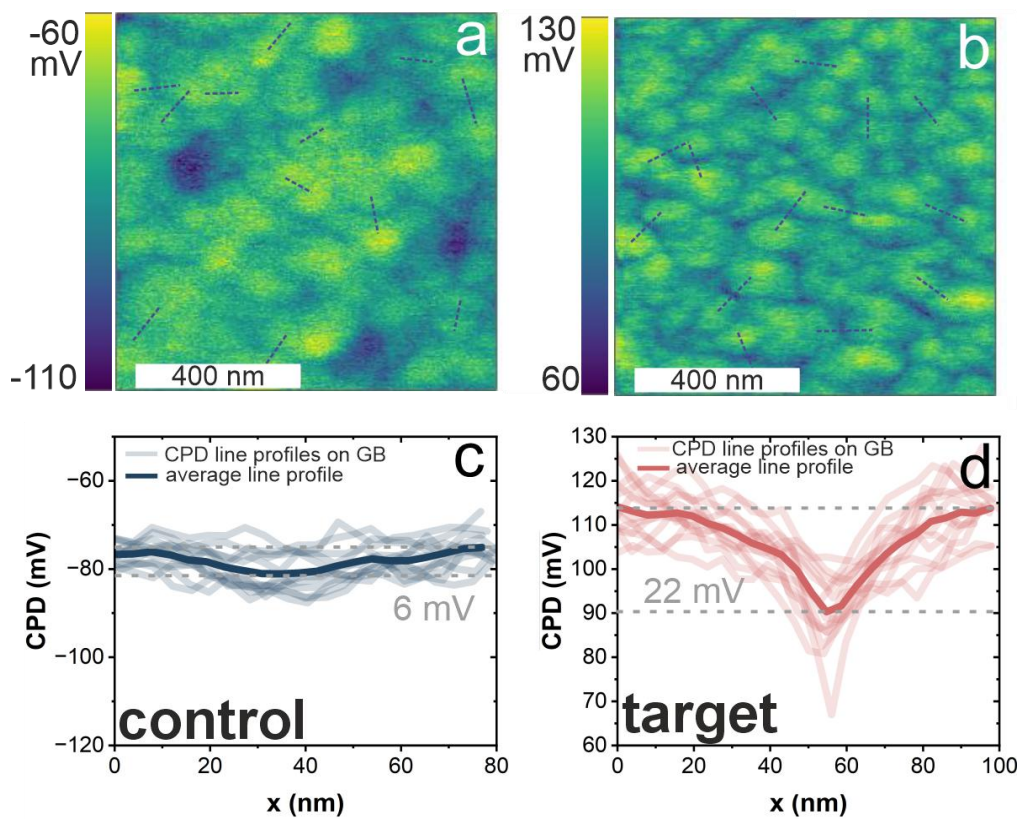

Figure 2 – Surface potential maps in control (a), and target (b) samples, and corresponding profiles (c), and (d) taken across black dashed lines

To estimate the impact of $AVA_2Pb_2I_7$ integration on the general optic properties in the fabricated PHS samples, we measured photoluminescence (**PL**) spectra and analyzed Tauc plots[35] (**fig. S2**, **ESI**). The intensity PL signal of Target thin-films was almost 45% higher compared to the Control one. Both sample configurations exhibited the symmetric PL peak at 778 nm (1.594 eV). The band-gap values, extracted from a Tauc plots of the absorption edge were 1.577 eV for Target and 1.565 for Control. The slight blue-shift (~0.01 eV) observed for PHS could be induced by intrinsic absorption of wide band-gap of $AVA_2FAPb_2I_7$ or originate from changes in lattice parameters, which could be discussed further. We analyzed the dynamic of radiative processes using time-resolved photoluminescence (**TRPL**) **(fig.S3**, **ESI**) and external quantum yield measurements (**PLEQE**) at different power densities (**fig.S4**, **ESI**). Time-resolved photoluminescence decay curves have been fitted with Shockley-Read-Hall model obtained from **eq.(1)**[36]. The extracted parameters presented in **tab. 1**.

$$n(t) = n_{0\_eh} \frac{\exp\left(-\frac{t}{tau_{SRH}}\right)}{1 + n_{0\_eh} \times beta \times tau_{SRH} \left(1 - \exp\left(-\frac{t}{tau_{SRH}}\right)\right)} + n_{0\_ex} \exp\left(-\frac{t}{tau_{EM}}\right) + c, \quad (1)$$

where $n_{0\_eh}$ is e-h weight, $n_{0\_ex}$ is exciton weight, $tau_{SRH}$ (ns) is Shockley-Read-Hall recombination lifetime, $tau_{EM}$ (ns) is excitonic emission lifetime, beta is bimolecular recombination rate, $c$ is noise level.

According to the fit, Target samples have higher $tau_{SRH}$ in two times than the control samples that could be associated with charge traps at interfaces or with an increase of a charge extraction. At the same time, the number of generated free carriers is higher for the target films, that is matched with PLEQE data presented in fig. S4 PLEQE of the target samples. The parameter $tau_{EM}$ for two untreated perovskite films is almost the same and rise of $n_{0\_eh}$ for the Target sample can be associated with an increase of photoluminescence external quantum yield. These optical data is the right evidence of an improvement of a charge generation in $AVA_2FAPb_2I_7$ - modified films. Structural imperfections and charge carrier traps are intrinsic to the crystallite surface. In microcrystalline $CsFAPbI_3$, SRH recombination is governed by states at both intergranular interfaces and film surfaces. The observed increase in photocarrier lifetime and quantum yield is attributed to the suppression of non-radiative recombination processes, facilitated by the impact of $AVA_2FAPb_2I_7$ on the interfaces of thin-films.

Table 1 – The extracted photoluminescence decay parameters For the PHS (Target) and reference $CsFAPbI_3$ (Control)

| Sample | $tau_{SRH}$ | $n_{0\_eh}$ | beta | $tau_{EM}$ | $n_{0\_ex}$ |
|---|---|---|---|---|---|
| control | 98 ± 1.5 | $9 \times 10^{12} \pm 7.6 \times 10^{12}$ | $0.0019 \pm 8 \times 10^{-5}$ | 324 ± 6.4 | 52 ± 0.6 |
| target | 213 ± 1.4 | $4 \times 10^{13} \pm 5.5 \times 10^{12}$ | $4.3 \times 10^{-4} \pm 1 \times 10^{-5}$ | 328 ± 11.5 | 44 ± 1.3 |

Modified films with AVAlPbI$_3$ shows PLEQE increase (see fig. S4) with excitation fluence till 0.5 W/cm$^2$. The target film shows a boost of the emission intensity after thermal tests, which can be associated with reduced traps concentration. The control perovskite film decreases PLEQE with growth of laser power density that can be associated with film instability under light fluence.

To investigate the structural differences between CsFAPbI$_3$ and PHS with AVA$_2$FAPb$_2$I$_7$, X-ray diffraction spectroscopy (**XRD, Fig. S5**, ESI) was employed. Both the control and target samples predominantly consist of the tetragonal β-CsFAPbI$_3$ phase, with characteristic peaks at 14.07°, 19.96°, 24.51°, 28.38°, and 31.81°, along with smaller peaks at 22.37° and 26.50°. In addition to the peaks corresponding to the ITO substrate at 21.30° and 30.34°, a peak corresponding to P-3m1 hexagonal lead iodide is observed at 12.66° for both samples. The unit cell parameters of the control sample, calculated from the diffractograms (**Tab. S1 in ESI**), are a = 8.909 Å and c = 6.308 Å. The formation of quasi-2D AVA$_2$FAPb$_2$I$_7$ increases the distance of the 110 plane from 6.310 Å to 6.353 Å, relative to the control perovskite's 110 crystal plane. The presence of AVA$_2$FAPb$_2$I$_7$ in the target film alters the lattice parameters to a = 8.901 Å and c = 6.302 Å (**tab.S1** in ESI). However, the low-angle XRD analysis (**Fig. S6**, ESI) of the Target film didn't reveal the expected peaks for the quasi-2D perovskite at 4.66° or 5.79°.

*Thin-films phase stability after thermal cycling*

Next, we conducted a study of the structural and interface stability of the reference CsFAPbI$_3$ and PHS with AVA$_2$FAPb$_2$I$_7$ under thermal cycling in ambient conditions (relative humidity (**RH**) was in the range from 30 to 60%). The exposure was performed in three regimes, with a lower temperature limit of -10°C and upper limits of +50, +80, and +100°C, respectively. We carried out thermal cycling (**TC**) at two-hour intervals (see schematic in **fig. S7**, ESI). The use of three specific modes for TC was based on the following rationale. FAPbI$_3$ owns metastable structural properties, which are characterized by spontaneous δ-phase (photo-inactive) nucleation occurring at grain boundaries or defects[37]. Environmental humidity, mechanical stress, and illumination could enhance this low-temperature phase instability, leading to degradation. Computational studies[38] suggest the energy barrier between δ- and α-FAPbI$_3$ phases is <0.1 eV, making room-temperature stabilization very complex. Partial substitution of FA$^+$ with Cs$^+$ in CsFAPbI$_3$ could reduce lattice strain and suppresses δ-phase formation by lowering the Gibbs free energy difference between phases[39]. Our reports show that using double cation CsFAPbI$_3$ provides relevant structural stability of the device stacks under constant heating with a thermostatic regime at 65°C[40]. However, for thermal cycling conditions, Cs-stabilized films exhibited gradual α-to-δ conversion due to repeated lattice expansion/contraction[41], [42]. Repeated cycling amplifies defect density at grain boundaries, creating pathways for moisture ingress and ion migration. TC under soft conditions (-10 to +50°C) potentially allows us to assess the contribution of spontaneous transitions to non-photoactive phases. Extending the upper temperature limit to 80°C and 100°C further enables the evaluation of decomposition mechanisms. Typically, decomposition is associated with thermally activated migration of Cs$^+$ and FA$^+$ ions, leading to the formation of insulating CsPbI$_3$ and FAPbI$_3$ domains. Simultaneously, oxidation processes contribute to degradation through the formation of I$_2$, Pb, and

PbI$_2$. Comparing phase composition changes across different thermal cycling regimes will help identify specific decomposition thresholds.

**Fig. S8** (**ESI**) displays the complete X-ray diffractograms obtained after TC across various temperature ranges. We assessed the phase stability of the thin-film samples by monitoring the intensity of the c β-FAPbI$_3$ perovskite peak, along with indicative markers of decomposition (PbI$_2$) and segregation (δ-FAPbI$_3$, δ-CsPbI$_3$). **Fig.3(a),(b)** illustrate the changes in the main peaks of β-FAPbI$_3$ (14.06°) and PbI$_2$ (12.66°) with an increase in maximum temperature of TC for the Control and Target samples, respectively. A noticeable increase in phase content of PbI$_2$ was observed for the Control samples after soft TC. However, the ratio of signal intensity for perovskite/PbI$_2$ phases remained stable (~0.5) as the upper TC limit was raised to 100°C *(**fig.3(c)**). By contrast, the Target samples displayed no indications of PbI$_2$-related decomposition, even for TC performed up to 100°C. An extra peak at 26.31° also appeared in Control perovskite films after soft TC up to max. 50°C. This can be assigned to the formation of the δ-FAPbI$_3$, δ-CsPbI$_3$, or the (211) reflection of β-CsFAPbI$_3$[43], [44]. So, the threshold for thermal decomposition and the α-to-δ phase transition in the Control CsFAPbI$_3$ thin-films was found to be near room temperature, while PHS didn't exhibit an evidence of phase segregation.

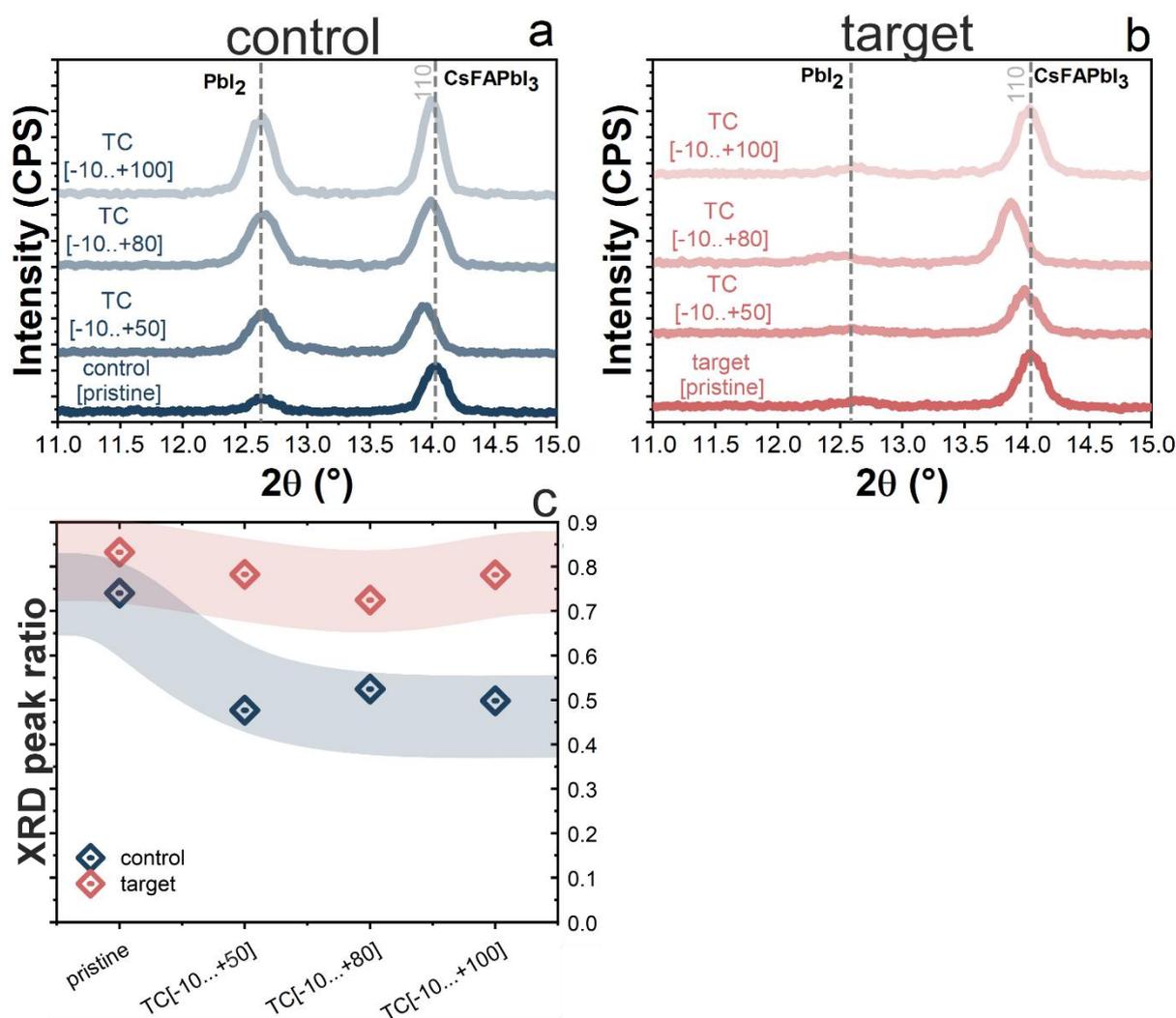

Figure 3 – Evolution of phase composition of Control (a) and Target (b) perovskite films after thermocycling in the climate chamber; relative integral intensity of 101 CsFAPbI$_3$ peak to 001 PbI$_2$ peak (c)

*Multi-layer stack phase stability after thermal cycling*

Beyond the phase stability of the photoactive layer in perovskite solar cells, chemically active hetero-boundaries could contribute to the degradation processes. The surfaces of halide perovskites exhibit significant dynamism due to their ionic nature[45], presence of the dangling bonds[46], [47] and point defects[48]. The application of an electric field can drive charge carriers toward the surface, accelerating electrochemical reactions occurring at the interface[49], [50]. This phenomenon can lead to changes in surface composition and structure, affecting both stability and optoelectronic properties. One of the most critical instability factors in p-i-n and n-i-p PSCs is oxidation of the back metal electrode. Ionic defects rapidly diffuse through charge-transport layers, reaching the electrode if the perovskite film surface is not passivated[51], [52], [53]. The accumulation of ions at the metal electrode results in corrosion and the formation of metal halides, which reduces the electrode's conductivity, induces non-radiative recombination. Actual literature reports[54], [55], [56], [57] partial stabilization of the back metal electrodes through multilayer metallization or integration of the diffusion barriers. However, a comprehensive solution to suppress the corrosion effects at interfaces requires modifying the intrinsic properties of perovskite films.

To investigate surface corrosion effects under an electric field at an unstable metal-perovskite interface, we analyzed a simplified ITO/perovskite/Cu multilayer structure. This sample configuration enabled the formation of an in-built electric field within the stack due to the work function difference between the electrodes and the perovskite layer, resulting in band bending. In parallel, we investigated glass/perovskite/Cu samples, where field effects were excluded, but the copper electrode remained in direct contact with the perovskite film.

To gather representative data, XRD analysis was performed on glass/perovskite/Cu structures subjected to thermal cycling between -10°C and +80°C (**Fig.4(a)**). For ITO/perovskite/Cu structures, XRD spectra were measured after "aging" process involving storage under ambient conditions ((**Fig.4(b)**): air exposure at 20–25°C and relative humidity (**RH**) ranging from 30% to 60% for 330 hours. Both Control and Target configurations of glass/perovskite/Cu stacks showed the signals of β-$FAPbI_3$ perovskite phase and presence of CuI released as a reaction product after TC. Also, the control sample showed phase segregation, producing $CsPbI_3$.

The aging of Control ITO/perovskite/Cu structures resulted in the formation of lead oxide, a notable amount of $CsPbI_3$, and δ-$FAPbI_3$. However, the target samples exhibited no significant changes in the phase composition. Visual inspection of ITO/perovskite/Cu structures showed a complete loss of absorption properties in the control sample after aging, specifically under the metal electrode. The corresponding absorption spectra for samples displayed in **fig.4(c)**. Localized perovskite degradation under the copper suggests that interfacial reactions between the perovskite and copper trigger phase changes in the photoactive layer. Perovskite/copper interaction mechanisms shifted when the interface of $CsFAPbI_3$ was modified by $AVA_2FAPb_2I_7$. In the target samples, the photoactive perovskite phase remained stable in the presence of copper, with $PbI_2$ appearing as a minor decomposition product after 300 hours of air exposure.

*Interface stability after thermal cycling*

Following an in-depth analysis of the structural changes in the multilayer stacks, we studied the chemical interactions at the perovskite/copper interface using the X-ray photoelectron spectroscopy (**XPS**). We considered the impact of oxygen and moisture in the ambient atmosphere to the surface of the thin-film stacks. The samples were kept in an inert environment and opened only immediately prior to loading into the vacuum chamber of the XPS bench. Survey XPS spectra of control and target films of ITO/perovskite/Cu (8nm) stacks are presented in **(Fig. S9 ESI)**. Representative Cu2p spectra for the control and target samples are shown in **fig.4(d,e).** The control sample exhibited several spectral changes, which were evaluated at multiple time points after aging in ambient conditions: 20 seconds, 10 minutes, 2 hours, 2 days and 6 days. By contrast, the target sample displayed much slower changes, with spectra recorded at the initial condition (20s) and even after 6 days of storage.

For the control samples, we observed in a Cu2p HR spectrum a component at 930.93 eV, corresponding to the overlap of the $I3p_{1/2}$ peak. Initially, the $Cu2p_{3/2}$ peak is observed at 932.56 eV, shifting to 932.67 eV in samples exposed to air for 10 minutes. After two hours and two days of exposure, the $Cu2p_{3/2}$ line was located at 932.43 eV and 932.45 eV, respectively. Notably, after 2 hours of storage, a component at 935.08 eV, attributed to Cu(II), begins to grow. This was accompanied by an increase in the intensity of the Cu(II) satellite peaks in the 946–940 eV range. After two days, the Cu(II) peak becomes comparable in intensity to the [Cu(0)+Cu(I)] peak, with a binding energy of 934.39 eV. The use of $AVA_2FAPb_2I_7$ in the target samples reduced the dynamics of copper oxidation on the perovskite surface (**Fig. 4(e)**). The main $Cu2p_{3/2}$ peaks were observed at binding energies of 932.33 eV at the start of the experiment and 932.37 eV after 6 days in air. Cu(II) components appear on the sample after 6 days of exposure but constitute less than 10% of all copper states in the sample. This fraction was calculated as the ratio of the satellite area to the total area of the main peak and satellite, as only Cu(II) contributes to satellite intensity.

Cu(II) components appeared in the sample after six days of exposure but accounted for less than 10% of the total copper states. This fraction was determined by calculating the ratio of the satellite area to the combined area of the main peak and satellite, as only Cu(II) contributes to satellite intensity[58]. The analysis of the Auger parameter is a highly accurate method for determining the state of copper thin-film. By combining the Cu $L_3M_{4,5}M_{4,5}$ line with the $Cu2p_{3/2}$ position, we made a Wagner diagram (**fig. 4(f)**) that shows good agreement in the values of the Auger parameter with reference data from the literature. Initially, the Auger parameter for both samples closely matches that of pure metallic copper (Cu(0)). Following air exposure, shifts in the Auger line indicated a chemical environment resembling CuI and $Cu_2O$. Notably, for the control sample, this evolution proceeds in two stages. In stage (i), Cu(0) transitions to a region associated with CuI, followed by stage (ii), where it further shifts to the $Cu_2O$ region. This behavior suggests a two-step oxidation process of copper on the perovskite surface under ambient exposure.

The stages of the oxidation process for the control sample were further confirmed by changes in the bonding states of perovskite components. High-resolution N1s spectra **(Fig. S10, ESI)** initially showed a single component at a binding energy of 400.43 eV for the control sample and 400.35 eV for the target sample. A second component with a binding energy of 399.04 eV emerges by the 2nd hour for the control sample and only after 6 days for the thin-films with $AVA_2FAPb_2I_7$. A third component at 397.59 eV is observed in the control sample after 2 days.

The component at approximately 400.5 eV could be assigned to the FA cation, while the nature of the additional components remains unclear. The copper-oxygen-iodine interaction proceeds with $FA^+$, involving intermediate steps in which the nitrogen binding energy transitions from 400.5 eV to 399.0 eV, and finally to 397.6 eV.

**Fig. S11** (ESI) illustrates the Pb4f spectra, which demonstrates an overall increase in peak intensity with extended air exposure, occurring alongside the thinning of the copper layer due to interactions with perovskite components. In the target sample, the $Pb4f_{7/2}$ peak area increased by 50% over 6 days, whereas the control sample showed a 40% increase. The oxidation state of lead in the target stack didn't change, as the binding energy remained stable at 138.28 eV till the end of the measurements. $Pb4f_{7/2}$ with a binding energy of 138.5 eV corresponds to $Pb^{2+}$ ions. In the control sample, however, the lead binding energy shifted to lower values, with the maximum shift occurring on the 2$^{nd}$ day of the storage. The $Pb4f_{7/2}$ peak position shifted from 138.49 eV at the start to 138.10 eV, representing a total shift of 0.4 eV.

Similarly, as for the lead spectra, $I3d_{5/2}$ for the target sample didn't shift (binding energy values of 619.04 and 619.02 eV) till the end of the measurement period (**fig. S12, ESI**). While for the control sample, the peak positions shift consistently to the right from 619.22 eV at the beginning of the experiment to 619.04 eV, 618.99 eV and 618.98 eV after 2 hours and two days, respectively. The Auger lines of the $I_{MNN}$ (**fig. 13, ESI**) for the control sample reveal a progressive shift of the main peak, transitioning from 968.51 eV to 968.28 eV, 968.22 eV, and ultimately 968.04 eV over a two-day period. A 72% decrease in the integrated intensity of the Auger peak accompanied this shift. For the Target stack, the initial Auger peak exhibits a pronounced doublet with binding energies of 968.36 eV and 966.79 eV. However, after 6 days, the shape of the $I_{MNN}$ peak for the target becomes identical to the control sample.

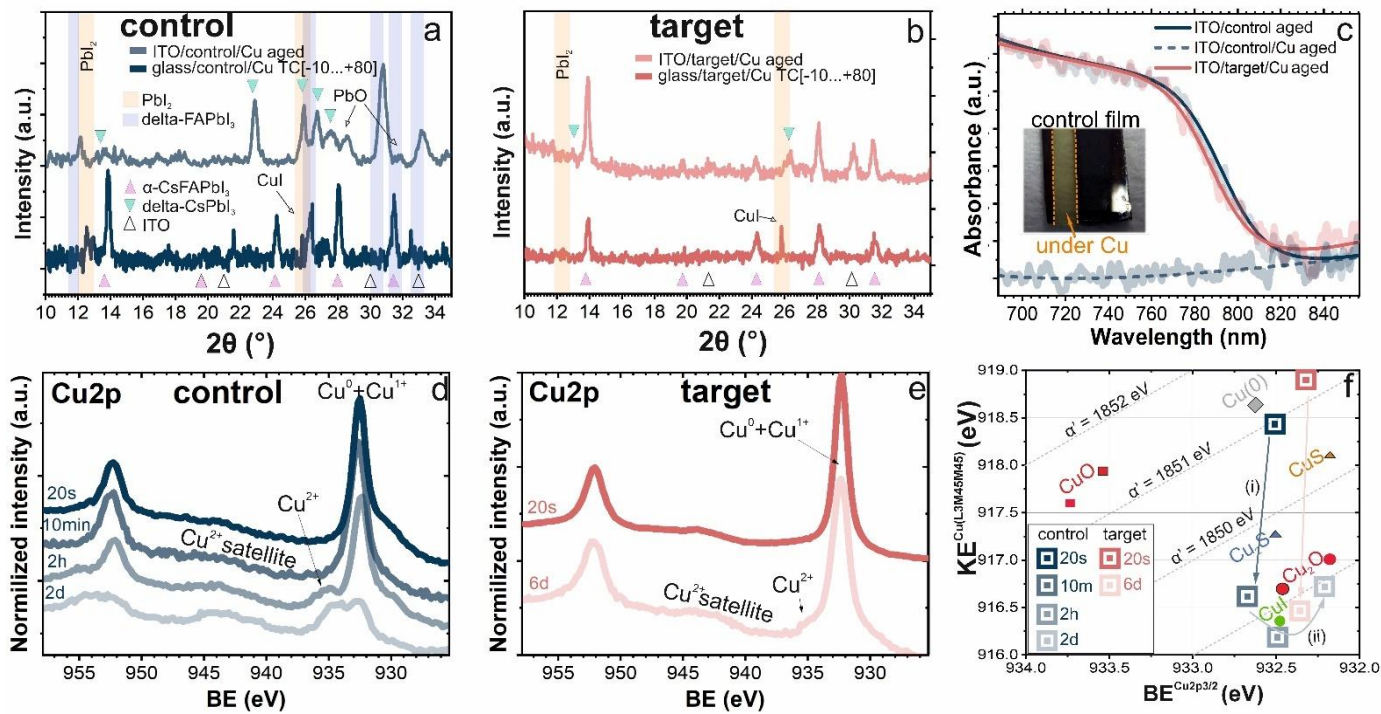

Figure 4 – X-Ray diffraction analysis of products perovskite and copper interaction in air with and without internal electric field for control (a) and target samples (b); optical absorption spectra of perovskite after 14 days in air (c); evolution of Cu2p spectra after reaction with perovskite in air for control (d) and target (e) samples; and (f) Cu $2p_{3/2}$ – Cu $L_3M_{4,5}M_{4,5}$ Wagner (chemical state) plot including literature data [58]

The obtained results allow us to discuss the mechanisms of corrosion processes at the perovskite/copper interface with the addition of $AVA_2FAPb_2I_7$. In the control sample, metallic copper decomposition progresses rapidly in several distinct stages upon air exposure. During the first stage, Cu(0) oxidizes to form a compound with an electron environment similar to CuI. Simultaneously, lead within the perovskite begins to reduce, partially transitioning from Pb(II) to Pb(0). Iodine also undergoes valence changes, likely forming crystalline iodine ($I_2$). Additionally, nitrogen appears to transform into a nitrogen-containing component.

After several days of air exposure, the chemical environment of copper shifts toward a state similar to $Cu_2O$, while the reduction of lead continues. The rate of iodine transformation slows, while a third nitrogen-containing substance is formed. The resulting products included CuI, Pb, and PbO. Certain compounds formed through the interaction of the organic copper cation with oxygen remain complex for precise determination.

Shifts in the position and changes in the shape of the N1s peak for perovskite/copper samples after ambient aging indicated an alteration in the states of the organic FAI cation ($CH(NH_2)_2I$), containing nitrogen. However, these spectral changes don't allow for precise identification of the new compounds formed during degradation. To further investigate the interaction products in the FAI-Cu system, we fabricated thin-film stacks, including ITO/FAI (200 nm) and ITO/FAI (200 nm)/Cu (100 nm) on the glass substrates. The samples were aged over 14 days under two conditions: ambient air with an average humidity of 35% and an inert atmosphere (dry N2 99.998%) inside a glove box. Comparative XRD and Fourier-transform Infrared spectroscopy (FTIR) of the samples before and after air exposure provided insight into the interaction products in this multi-component system. A comparison of samples stored in an inert and air atmosphere helped to determine the role of oxygen and moisture in interaction processes in the FAI/Cu system. Analysis via XRD

(**Fig.S14 (a),** ESI) demonstrated that formamidine molecules are stable and don't decompose during aging in air or upon GB exposure. We indicated the corresponding peaks of FAI for the samples exposed to the air and for the multi-layers with copper (stored in inert atmosphere) at 18.00°, 24.55°, 25.73°, and 27.01° [59]. However, after exposure of ITO/FAI/Cu stacks under ambient conditions, the intensity of the diffraction signals decreased. FTIR spectral changes revealed the results of decomposition and interaction in the FAI-Cu system (**Fig. S14 (b)**, ESI). Peaks representing oxidized carbon (C=O at 1709.4 cm-1) and nitrogen (N-O at 1546.4 cm-1) within the 1800–1500 cm$^{-1}$ range appeared only for the air-aged FAI/Cu sample (**Fig. S14 (c)**, ESI). Furthermore, FAI oxidation produced C-O bonds, identified through specific wave-numbers (1360.2, 1293.6, and 1184.6 cm$^{-1}$, **Fig. S14 (d)**, ESI). Thus, the interaction between FAI and copper occurs only in the presence of moisture and oxygen.

In the perovskite molecule, organic cations inhibit the transition of PbI$_6$ octahedrons to thermodynamically stabilized δ - phase. During the reaction with copper, the FA-cation on the perovskite surface is consumed. This leads to the compression of lead iodide octahedrons, causing a phase transition to the δ-phase. In regions with a local deficiency of the FA-cations, PbI$_2$ formation is possible. When a critical accumulation of the delta phase is achieved, it induces the rearrangement of neighboring perovskite atoms due to its higher energetic favorability. The presence of an electric field may facilitate the accumulation of positively charged FA$^+$ ions at the copper/perovskite interface, accelerating the decomposition of the perovskite.

Our results showed that, in the target thin-films, the oxidation states of iodine and lead remain largely unchanged. Although the formation of Cu(II) goes slowly, the Auger parameter α' reveals that copper transitions to a partially oxidized state (Cu(0) + Cu(I) + Cu(II)) after 6 days of aging in ambient conditions. Nitrogen was involved in the reactions, but didn't produce compounds similar to those observed in the control sample. Thus, the corrosion mechanisms of the copper electrode differ between the CsFAPbI$_3$ and samples with AVA$_2$FAPb$_2$I$_7$. For the target thin-films, lead and iodine reduction products from the perovskite don't participate in reactions with copper, resulting in a marked improvement in chemical stability at the perovskite/copper interface. Control stacks of ITO/perovskite/Cu exhibited a rapid phase transition from the photoactive perovskite α or β phase to the yellow δ phase under an electric field and air exposure. This process was initiated within the first two hours of air exposure. Subsequently, corrosion processes accelerated within minutes, with the recrystallization front becoming visible. The stages of copper oxidation on the perovskite surface can be attributed to the different reactive capability of the α and δ phases of halide perovskite.

In general, the following mechanisms can explain the stabilization of the perovskite/copper interface upon the addition of AVA$_2$FAPb$_2$I$_7$. First, the surface interlayer of AVA$_2$FAPb$_2$I$_7$ act as a restraining agent, increasing the activation energy required for the phase transition from the photoactive phase. Second, furthermore, the ionic diffusion to electrode interface is mitigated by AVA$_2$FAPb$_2$I$_7$, which fills intergranular voids, thus decreasing ion migration within the perovskite volume under internal electric fields.

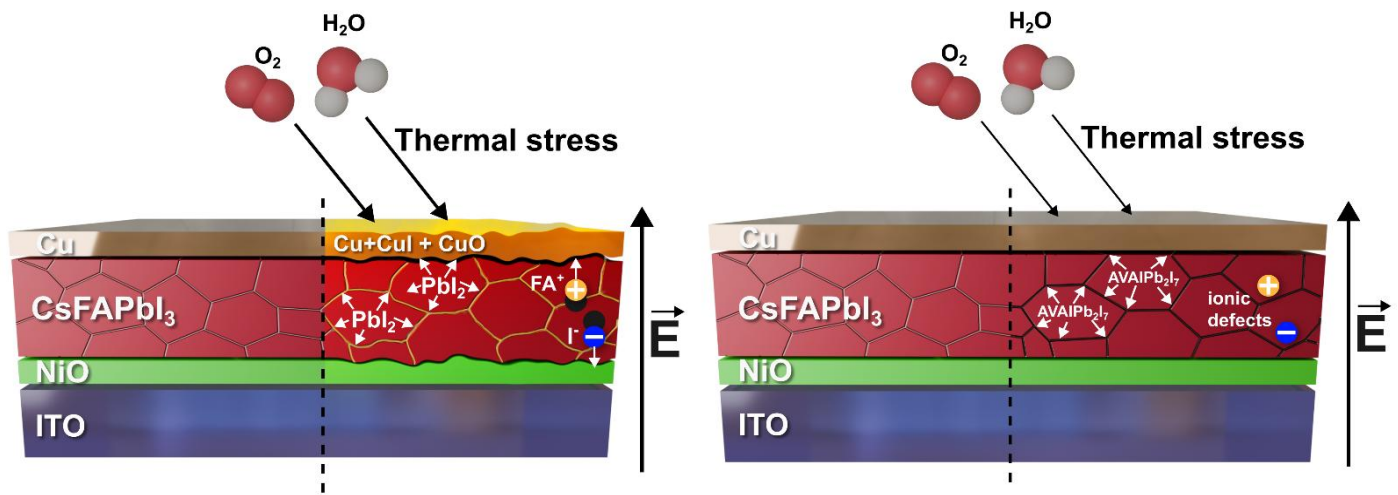

Figure 5 - Representative schematics of the interface stabilization in microcrystalline $CsFAPbI_3$ with $AVA_2FAPb_2I_7$

Current state-of-the-art (SoA) applications of AVA-based organic cations focus on stabilizing various perovskites and elucidating the mechanisms that mitigate corrosion effects, primarily for single-cation compositions. For $MAPbI_3$, previous studies have reported hetero-boundary modification and reversible decomposition effects[60]. Recently, *Perrin et al.*[61] demonstrated that AVAI additives effectively suppress bulk degradation in thin-film $MAPbI_3$, although interface corrosion remained a critical issue. Alnzai et al.[62] conducted an in-depth investigation into the stabilization of $FAPbI_3$, with NMR analysis revealing the formation of hydrogen-rich bonds contributing to surface passivation. Our work provides key insights into the stabilization of double cation formulation – $CsFAPbI_3$, which are widely utilized in high-performance single-junction and tandem solar cells. The results demonstrate a combination of enhanced phase stability, suppression of ionic defect migration, and improved stabilization of the perovskite/metal interface under ambient conditions. This approach holds significant potential for mitigating corrosion effects in large-scale perovskite modules, particularly at the perovskite/back electrode interface following P3 structuring (isolation of the sub-cell after metallization[63]).

**Conclusions**

The integration of $AVA_2FAPb_2I_7$ for the modification of $CsFAPbI_3$ directly altered the inter-grain energetics and enhanced chemical stability under various external factors – thermal cycling, humidity, electric field. Enhanced phase resilience in perovskite heterostructures, even after 100°C heating, resulted from increased energy activation for δ- phases formation and decomposition at interfaces. Using $AVA_2FAPb_2I_7$ induced compensation of the non-radiative defects, mitigated the ion migration and promoted passivation of the $CsFAPbI_3$ surface. Specifically, the $AVA_2FAPb_2I_7$ stabilized the states of iodine, lead and nitrogen, which inhibited rapid oxidation processes at the metal/perovskite interface triggered by accumulation of the ionic species, and presence of humid and oxygen. Complex surface studies for the interaction with copper contacts revealed stable Cu(0)/Cu(I) states over prolonged exposure, whereas the unmodified $CsFAPbI_3$ quickly evolves to higher copper oxidation states, accompanied by lead reduction and the formation of nitrogen-related degradation products. So, observed findings demonstrated that forming an intergranular barrier, $AVA_2FAPb_2I_7$ reduces both the nucleation of unfavorable phases and the diffusion of reactive species toward metal electrodes. The obtained results output important specifics for the intrinsic stabilization of the microcrystalline perovskite absorbers with improved durability under thermal stress and applied electric fields in ambient conditions.

## Notes

The authors declare no competing financial interest.

## Supporting Information

The electronic supplementary material contains the following data:

## Acknowledgements

The authors gratefully acknowledge the financial support from the Russian Science Foundation (RSF) with grant № 24-62-00022.

**Electronic supplementary information for the manuscript:**

**"Stabilization of interfaces for double-cation halide perovskites with AVA$_2$FAPb$_2$I$_7$ additives"**

**By**

Lev O. Luchnikov[1], Ekaterina A. Ilicheva[1], Victor A. Voronov[1], Prokhor A. Alekseev[2,3], Mikhail S. Dunaevskiy[2,3], Vladislav Kalinichenko[2], Vladimir Ivanov[2], Aleksandra Furasova[2,4], Daria A. Krupanova[5], Ekaterina V. Tekshina[5], Sergey A. Kozyukhin[5], Dmitry S. Muratov[6], Maria I. Voronova[1], Danila S. Saranin[1,2#], and Eugene I. Terukov[2]

**Affiliations:**

[1]LASE – Laboratory of Advanced Solar Energy, NUST MISiS, 119049 Moscow, Russia

[2]School of Physics and Engineering, ITMO University, 197101, St. Petersburg, Russia

[3]Ioffe Institute, 194021, St. Petersburg, Russia

[4]Qingdao Innovation and Development Center, Harbin Engineering University, Qingdao 266000, Shandong, China

[5]Kurnakov Institute of General and Inorganic Chemistry of the Russian Academy of Sciences, Moscow, 119991, Russia

[6]Department of Chemistry, University of Turin, 10125, Turin, Italy

**Experimental section:**

**Ball milling**

To prepare AVA$_2$FAPb$_2$I$_7$ Powder precursors of 5-aminovaleric acid iodide(5-AVAI), lead (II) Iodide (PbI$_2$) and Formamidine iodide (FAI; ultra-dry, 99.999%, Lanhit) in molar relation 2:2:1 (AVAI:PbI$_2$:FAI) were prepared in grinding jars with ZrO$_2$ beads under nitrogen atmosphere. Precursors were mixed in ball milling machine by multi-axes rotation (Spex Sample Prep 8000M) in 8 cycles. Each cycle took 10 minutes rotation 800 revolutions per minute and 10 minutes break for heat dissipation. After the milling process mixed powders were stored in nitrogen.

**Solution preparation**

For the preparation of 1 ml of perovskite ink, powders with the following mass values were mixed: FACl (9,66 mg), CsI (62,35 mg), FAI (144,45 mg), PbI$_2$ (553,2 mg). The resulting mixture was dissolved in a DMF:NMP (volume ratio 9:1) and stirred at a temperature of 50 °C for 1 h. After complete dissolution, the perovskite solution was filtered and placed in a vial with mechanosynthesized AVA$_2$FAPbI$_7$ powder (15,64 mg) and stirred for 10 minutes.

**Films deposition**

Firstly, the ITO substrates were cleaned with detergent, de-ionized water, acetone, and IPA in the ultrasonic bath. Then, substrates were activated under UV-ozone irradiation for 30 min. NiCl$_2$·6H$_2$O+HNO$_3$ precursor for NiO HTM film was spin-coated at 4000 RPMs (30 s), dried at 120 °C (10 min), and annealed at 300 °C (1 h) in the ambient atmosphere. The Perovskite absorber film was crystallized on the top of NiO$_x$ with spin-VASP method. The Perovskite precursor was spin-coated at 2000 RPMs (4 s). Then, substrates were placed in a VASP chamber for 3 minutes under vacuum. After vacuum treatment, the substrates were placed on a plate for annealing at 105 °C (30 min) for conversation into the black perovskite phase. The copper cathode was deposited with the thermal evaporation method at $2 \times 10^{-6}$ Torr vacuum level through a shadow mask.

**Characterization**

X-Ray diffraction **(XRD)** of perovskite layers was investigated with diffractometer Tongda TDM-10 using CuKα as a source with wavelength 1.5409 Å under 30 kV voltage and a current of 20 mA. absorbance spectra **(ABS)** of perovskite films was carried out via SE2030-010-DUVN spectrophotometer with a wavelength range of 200–1100 nm. Photoluminescence spectra **(PL)** were recorded on Agilent Cary Eclipse Fluorescence Spectrophotometer on 550nm excitation wavelength.
AFM and KPFM measurements were performed in room ambient conditions using Ntegra AURA (NT-MDT) microscope. NSG10/Pt (Tipsnano) probes were used with tip curvature radius 30 nm. KPFM study utilized an amplitude modulation regime. AFM and KPFM images were obtained simultaneously by two-pass method.

Time resolved photoluminescence decay **(TRPL)** of the perovskite films was performed with use of time correlated photon counting setup. For these studies, PL signal was excited at 532 nm wavelength with a laser diode having the pulse duration of 100 ps and 250 kHz repetition rate, while the photon counting detector modules with a timing resolution of 50 ps was used to detect luminescence decay. The configuration for TRPL signals has been measured by reflection with use of NIR objective 10X/NA 0.26. The excitation beam was unfocused, and the laser spot had diameter approximately 160 μm. ATR spectra were measured with use of Shimadzu UV-3600 Plus in 400 - 2000 nm range. Photoluminescence quantum yield measurements of the perovskite films were performed using 405 nm CW laser, Ocean Optics QE Pro spectrometer and Labsphere Integrated Sphere. LabView program was used to calculate **(PLQY)** values from PL spectra.

The Fourier-transform infrared **(FTIR)** spectroscopy measurements were conducted using a Simex FT-801 IR spectrometer equipped with an attenuated total reflectance (ATR) accessory. X-ray photoelectron spectroscopy **(XPS)** analysis was carried out with a PREVAC EA15 electron spectrometer. The primary radiation source utilized was AlKα (hν = 1486.6 eV, 150 W). Calibration of the binding energy (BE) scale was performed using the characteristic binding energies of Ag3d5/2 (368.3 eV) and Au4f7/2 (84.0 eV) from silver and gold foils, respectively. Fresh Perovskite film samples were kept in a nitrogen atmosphere prior to measurement. Exposure to air lasted less than 20 seconds before the samples were transferred to the vacuum chamber.

REOCAM TCH-150 climatic chamber was used to conduct climatic tests with cyclic changes in temperature and humidity. Tests were carried out without additional lighting. The usable volume of the chamber is 159 liters. The amplitude of temperature fluctuations in the steady-state mode is 0.5°C. Ttemperature ununiformity over the volume in the steady-state thermal mode was not more than 4°C. Time transition time from one temperature point to another is 6 min. Cooling method cooling method - air cooling with the help of two-cascade cooling system based on Tecumseh (Embraco) compressors.

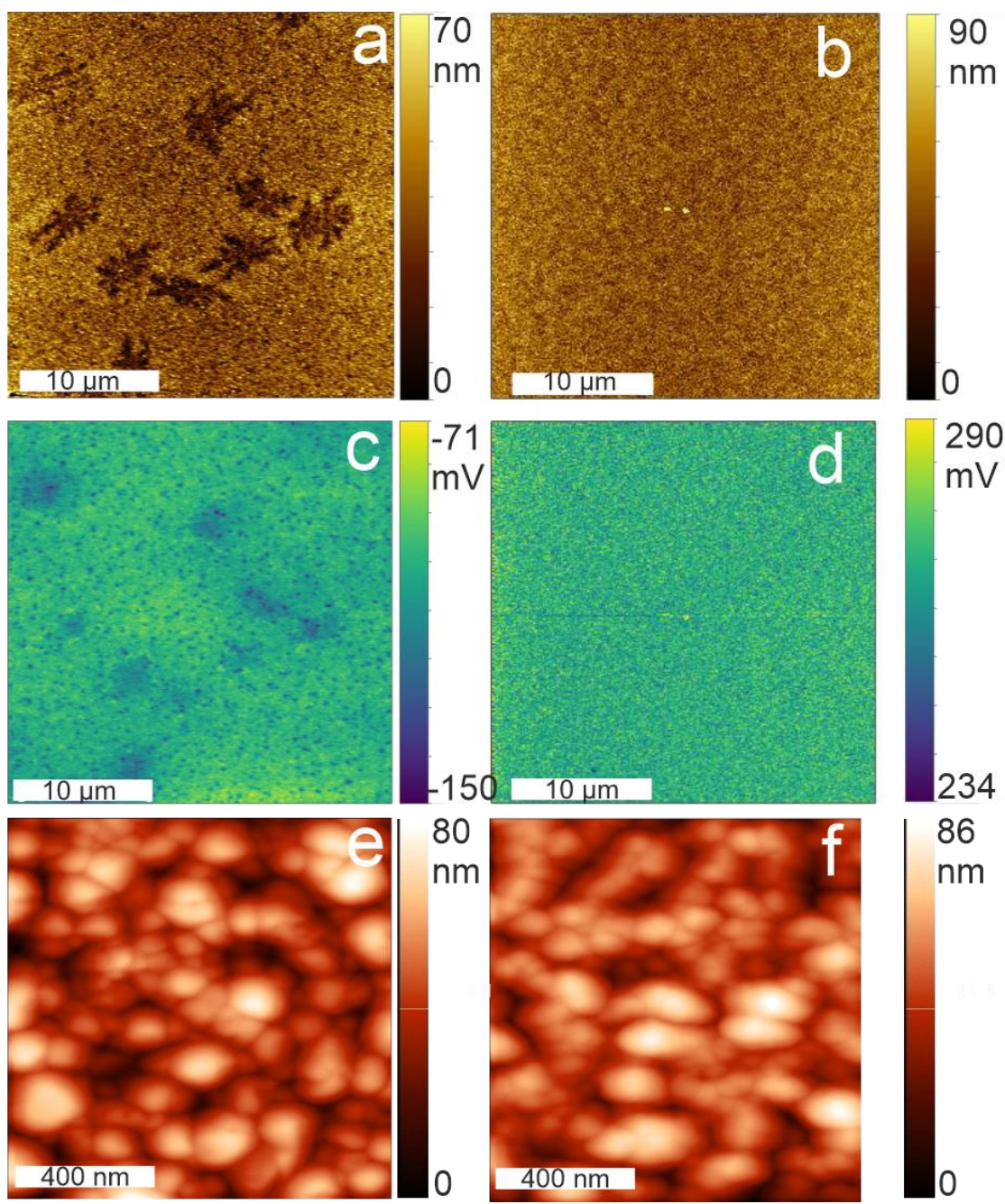

Figure S1– topography of control (a) and target (b) perovskite films, surface potential of control (c) and target (d) and grains morphology of control (e) and target (f) samples

*Optical characterization*

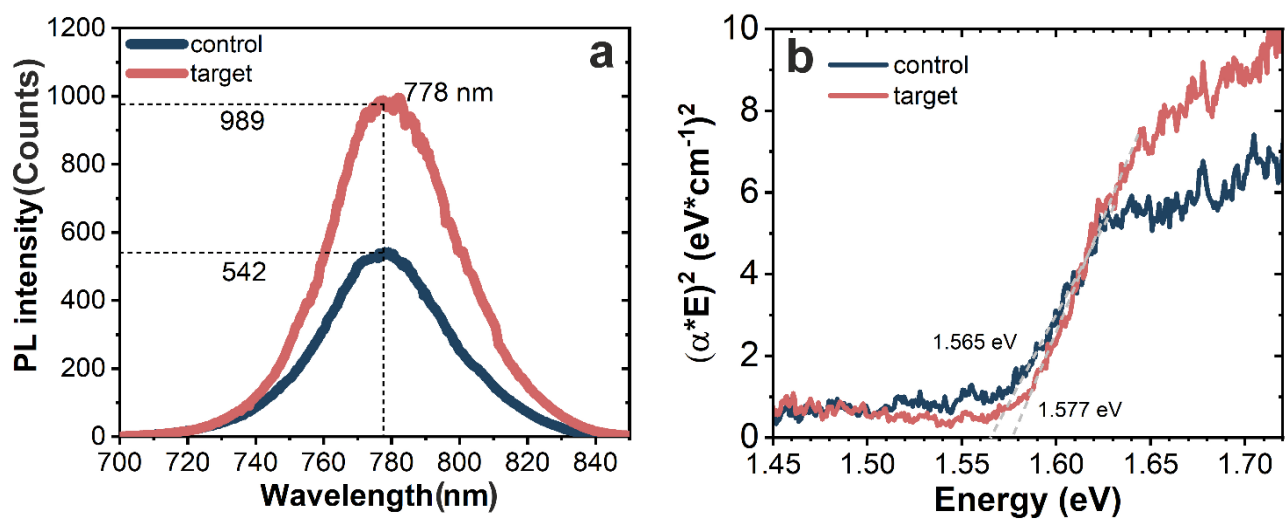

Figure S2 – PL spectra for the Control and Target samples (a); Tauc plots for the optical absorption (b)

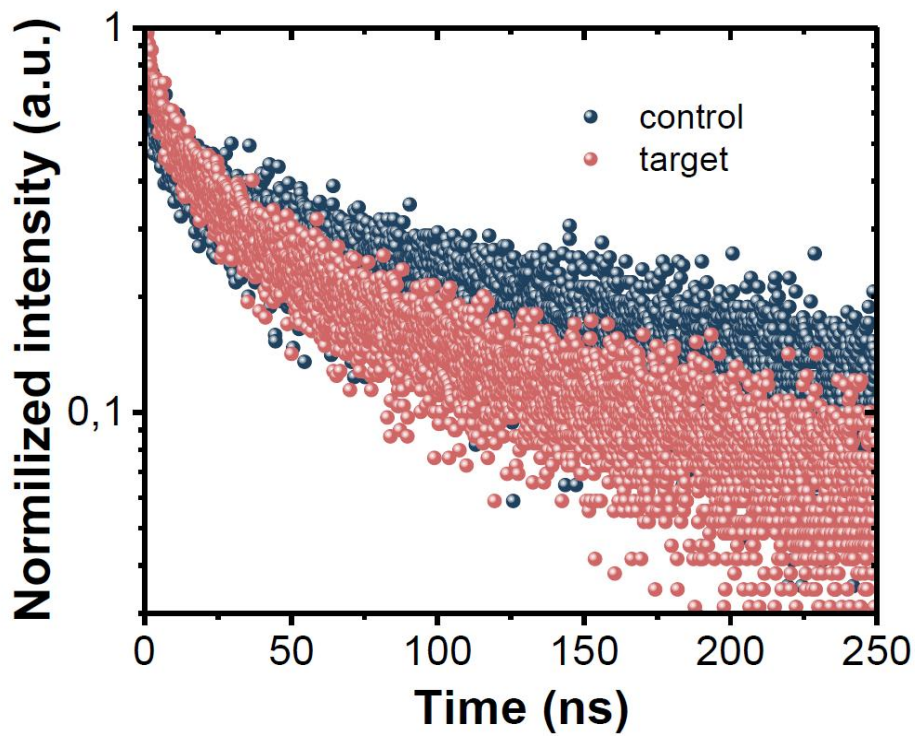

Figure S3 -– Time resolved photoluminescence for CsFAPbI$_3$ (control, dark blue data) and CsFAPbI$_3$ modified with AVA$_2$FAP$_2$bI$_7$ (target, coral data)

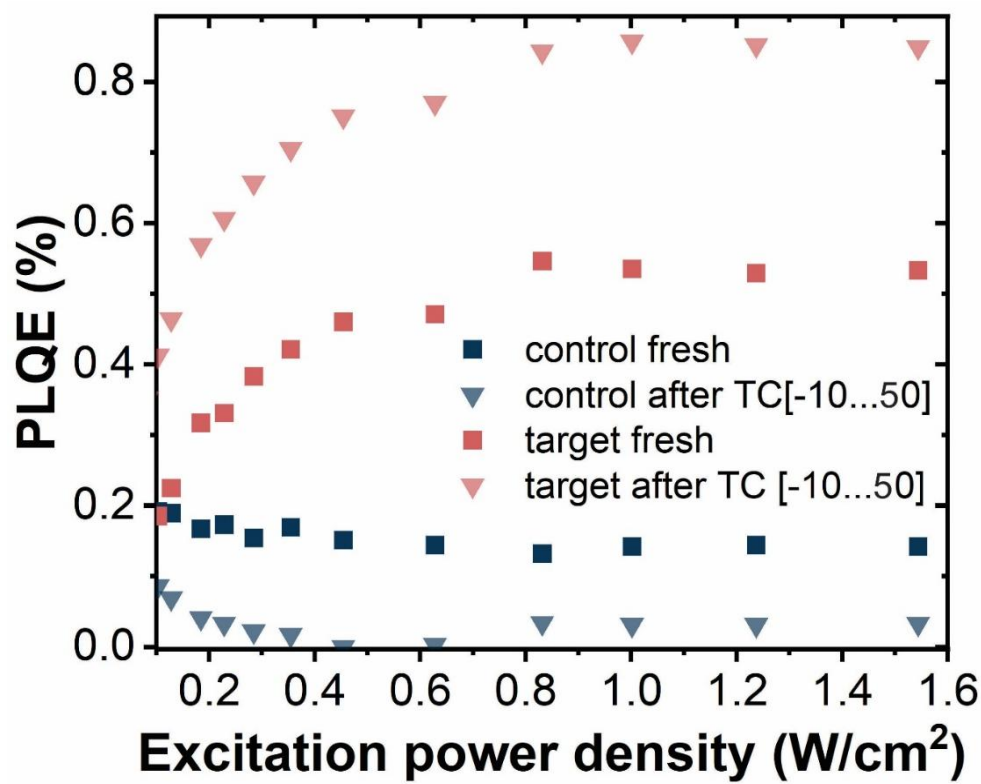

Figure S4– photoluminescence external quantum yield under different excitation power energy for untreated CsFAPbI$_3$ (control, dark blue data), CsFAPbI$_3$ modified with AVA$_2$FAP$_2$bI$_7$ (target, coral data) and after thermal cycling test for CsFAPbI$_3$ (control, light blue data), CsFAPbI$_3$ modified with AVA$_2$FAP$_2$bI$_7$ (target, pink data).

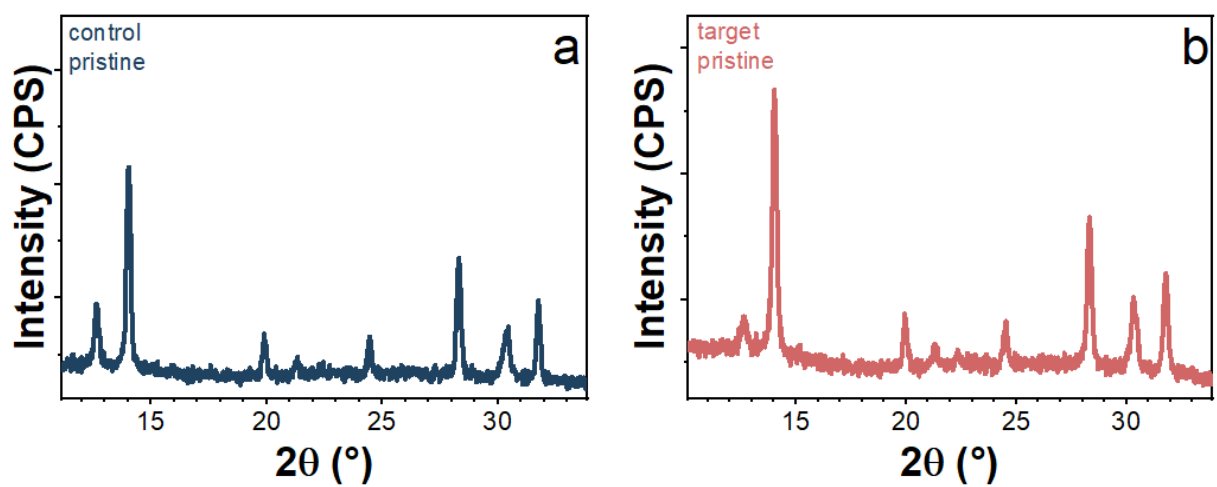

Figure S5 – XRD of pristine control (a) and target (b) perovskite films

Table S1 – The calculated lattice parameters of Control and Target thin-films obtained with Rietveld method

| lattice parameter | control | target |
|---|---|---|
| a | 8.888 Å | 8.879 Å |
| c | 6.288 Å | 6.297 Å |

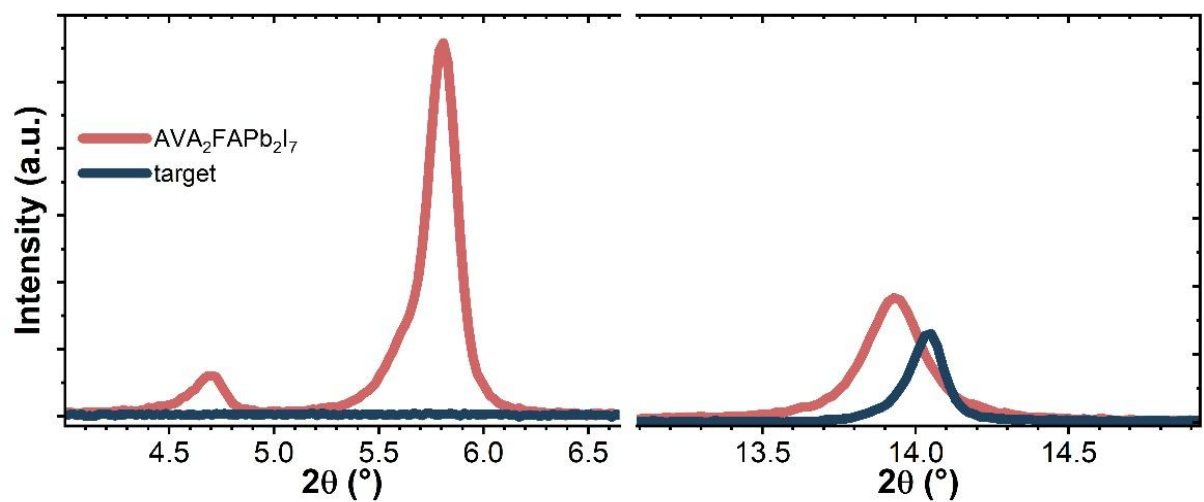

Figure S6 – comparison of X-ray diffraction of target film with $AVA_2FAP_2bI_7$

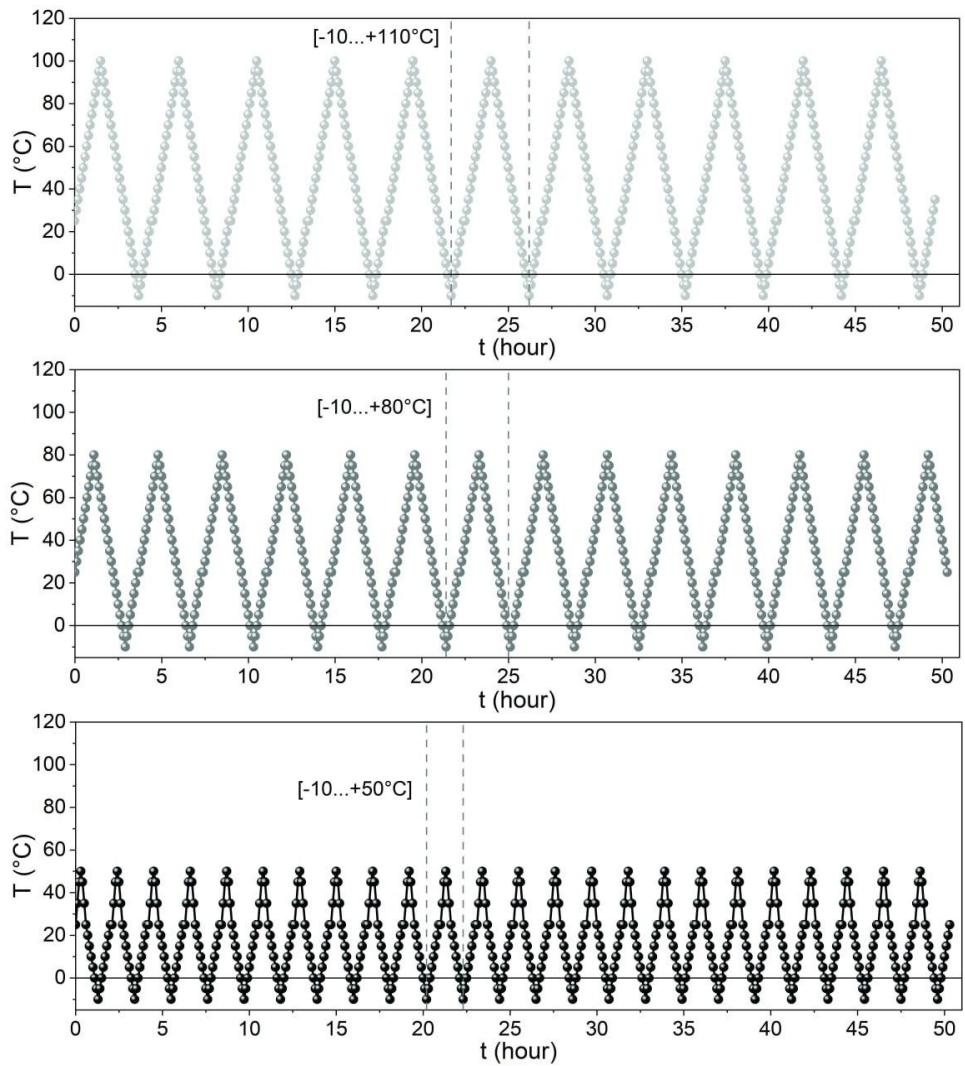

Figure S7 – The plots of corresponding thermal cycling regimes used for the investigation

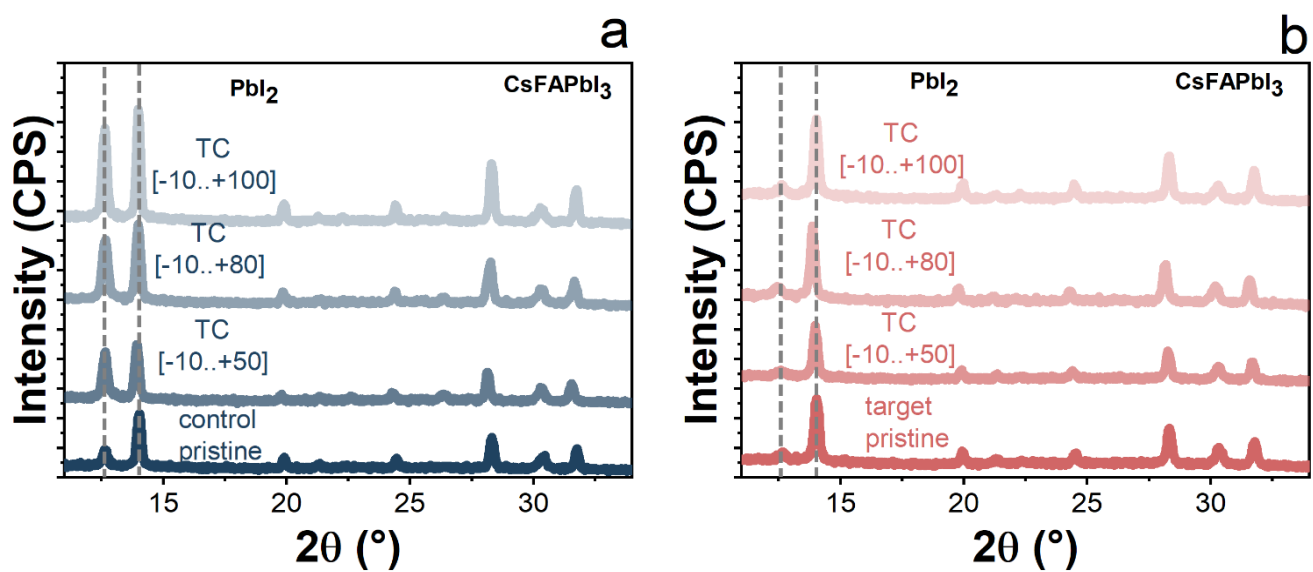

Figure S8 – Evolution of phase composition of Control (a) and Target (b) perovskite films after thermocycling in the climate chamber. The ITO substrate showed diffraction peaks at 21.13° and 30.15°

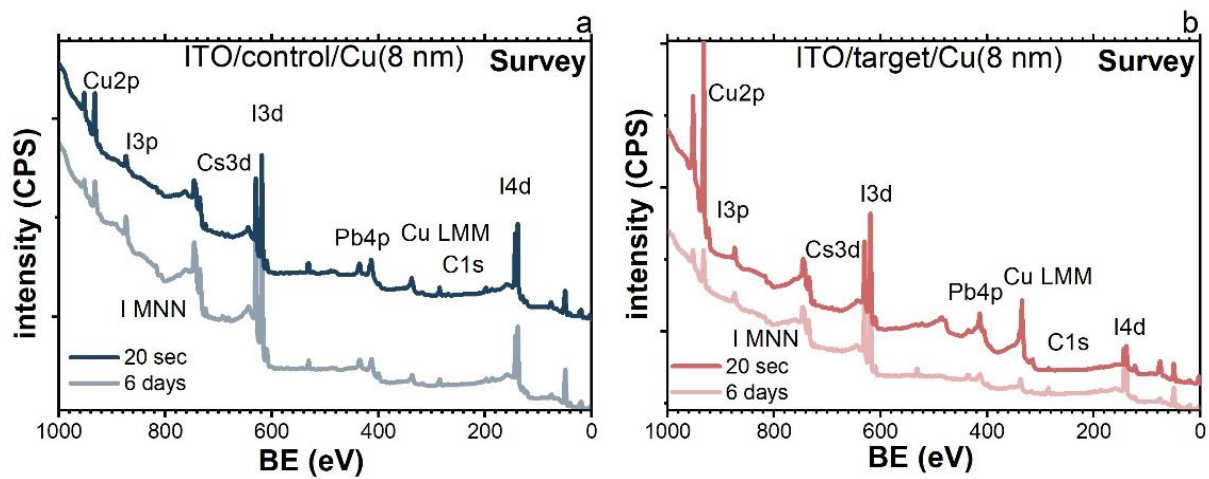

Figure S9 – Survey spectra of control (a) and target (b) perovskite films in multilayer stack with Cu before and after agin.

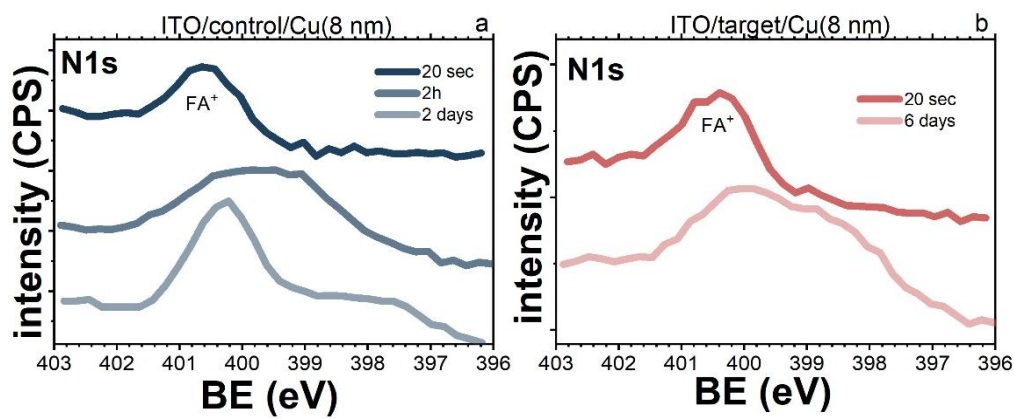

Figure S10 – Evolution of HR XPS N1s spectra of control (a) and target (b) samples with Cu

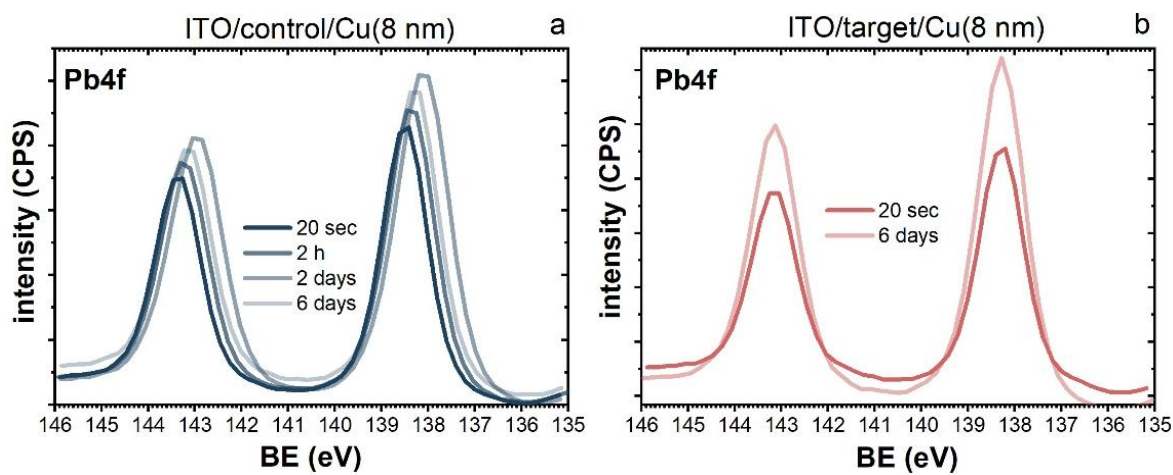

Figure S11 – Evolution of HR XPS Pb4f spectra of control (a) and target (b) samples with Cu

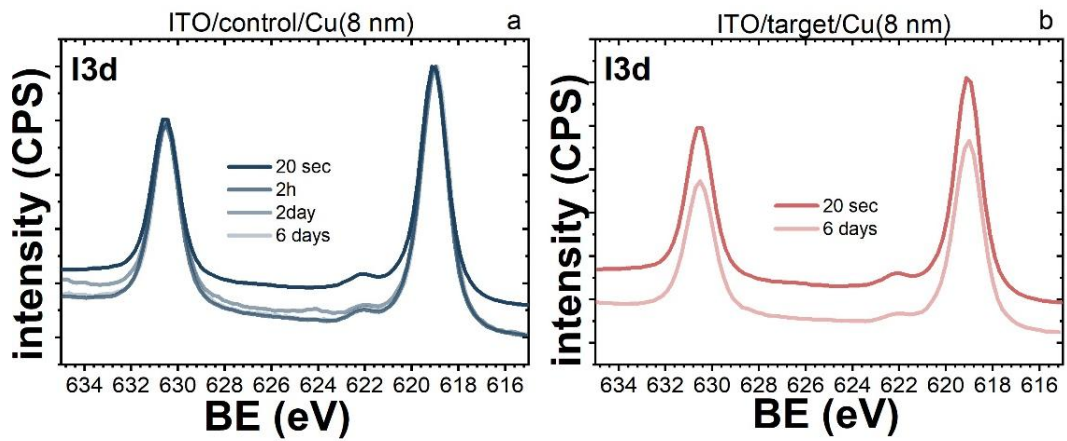

Figure S12 – Evolution of HR XPS I3d spectra of control (a) and target (b) samples with Cu

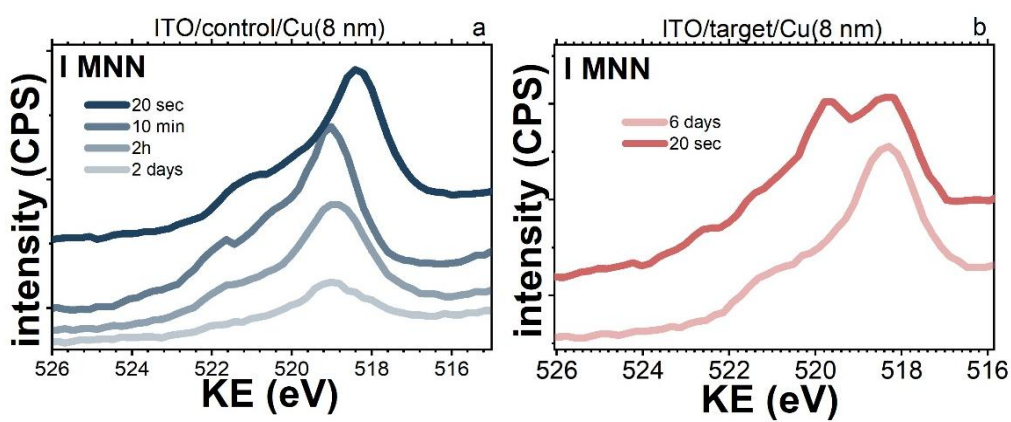

Figure S13 – Evolution of HR XPS $I_{MNN}$ spectra of control (a) and target (b) samples with Cu

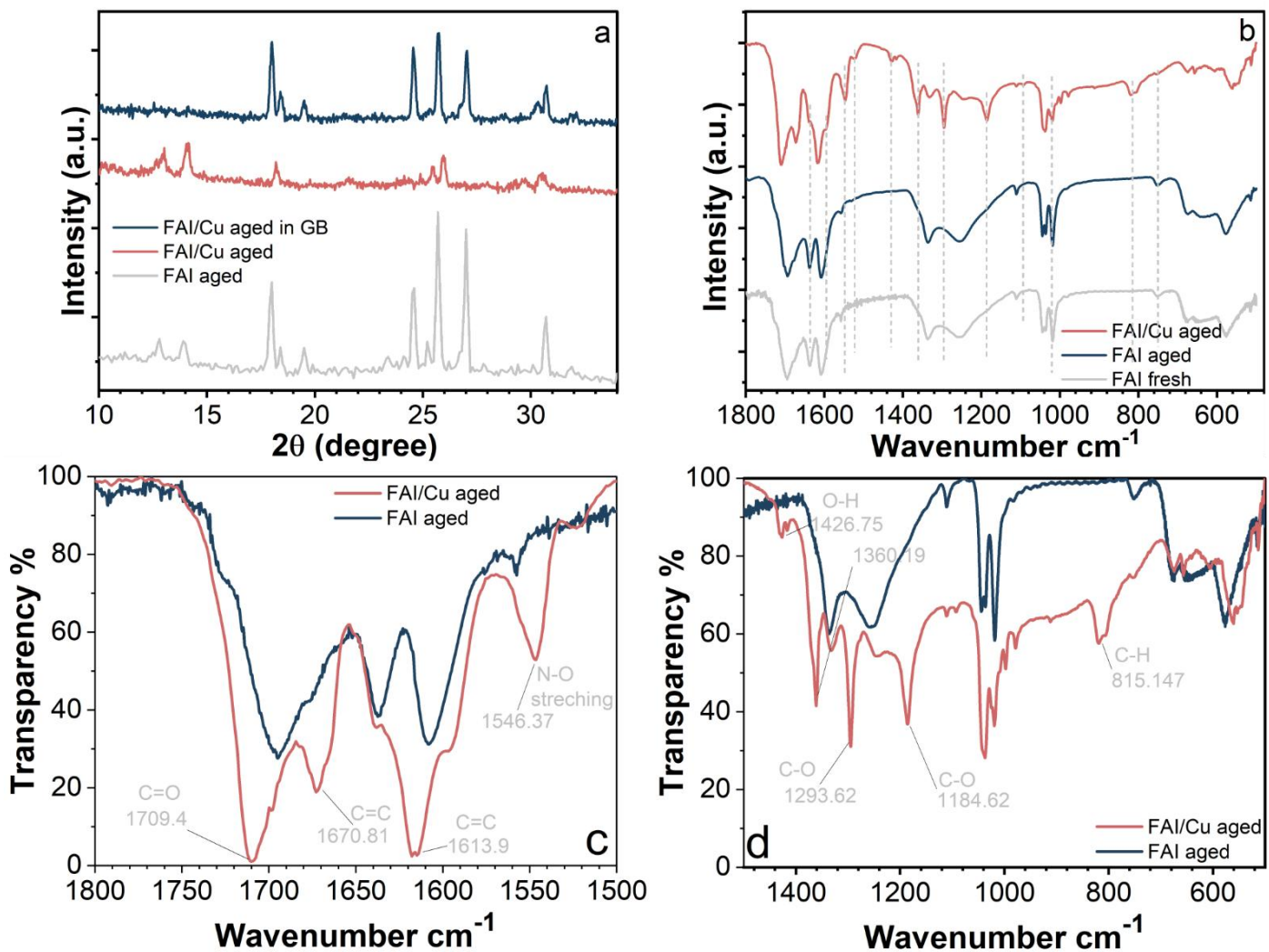

Figure S14 – XRD (a). and FTIR (a-d) analysis of FAI/Cu and FAI samples on ITO substrate after agin